\documentclass[%
preprint,
amsmath,amssymb,
aps,
]{revtex4-2}

\usepackage{graphicx}
\usepackage{dcolumn}
\usepackage{bm}
\usepackage{gensymb}
\usepackage{textcomp}
\usepackage{todonotes}
\usepackage[colorlinks=true,citecolor=blue,linkcolor=blue]{hyperref}

\usepackage[normalem]{ulem}

\usepackage{cancel}
\usepackage{subeqnarray}
\usepackage[mathcal]{euscript}
\def\mathbi#1{\textbf{\em #1}}
\newcommand{\h}{\mathcal{H}}
\def\be{\begin{equation}}
\def\ee{\end{equation}}

\def\pder#1#2{\frac{\partial #1}{\partial #2}}
\def\md{\mathrm{d}}
\def\oder#1#2{\frac{\md #1}{\md #2}}
\newcommand{\gt}{\tilde{g}}
\def\dd{\mathrm{d}}
\newcommand\norm[1]{\left\lVert#1\right\rVert}

\usepackage{verbatim}

\usepackage{cleveref}

\crefformat{section}{\S#2#1#3} 
\crefformat{subsection}{\S#2#1#3}
\crefformat{subsubsection}{\S#2#1#3}

\def\dd{{\rm d}}
\def\ee{{\rm e}}

\begin{document}

\preprint{APS/123-QED}

\title{The motion of buoyant point vortices}

\author{Jeffrey R. Carpenter}
\affiliation{Institute of Coastal Research, Helmholtz-Zentrum Geesthacht, Geesthacht 21502, Germany.}

\author{Anirban Guha}%
 \email{anirbanguha.ubc@gmail.com}
\affiliation{School of Science and Engineering, University of Dundee, Nethergate, DD1 4HN, U.K.}





\date{\today}

\begin{abstract}
    A general formulation is presented for studying the motion of buoyant vortices {in a homogeneous ambient fluid.}  It extends the well-known Hamiltonian framework for interacting homogeneous point vortices to include buoyancy effects acting on the vortices.  This is then used to systematically examine the buoyant 1-, 2-, and 3-vortex problems.  In doing so we find that 2 buoyant vortices may either evolve as a pair in bounded circular orbits, or as two independent unbounded vortices that drift apart, and a criteria is found to distinguish these cases.  Special attention is given to the buoyant vortex couple, consisting of two vortices of equal and opposite circulation, and equal buoyancy anomaly.  We show that a theoretical maximum height is generally possible for the rise (or fall) of such couples against buoyancy forces.  Finally, the possibility and onset of chaotic motions in the buoyant 3-vortex problem is addressed.  In contrast to the homogeneous 3-vortex problem, the buoyant vortex system shows evidence that chaos is present.  We also demonstrate the chaotic advection of tracer parcels arising from the flow field induced by just 2 buoyant vortices.

\end{abstract}

\maketitle


\section{Introduction}


Ever since the classical paper of \citet{helm1858}, vortex dynamics has been used as a tool to understand the behaviour of complex fluid flows.  In the words of \citet{aref1983}: `...the evolution of vorticity, and thus the motions of vortices, are essential ingredients of virtually any real flow.  Hence vortex dynamics is of profound practical importance'.  Based on this principle, in the present paper we consider a general model for the interaction of singular `point' vortices that are subject to buoyant forces {in a homogeneous ambient fluid}.  We find that the presence of buoyancy leads to richer dynamics than in the homogeneous vortex problem, and displays some non-intuitive solutions,  including the more rapid appearance of chaos compared to homogeneous flows as the number of vortices is increased.  The present work generalises previous studies to account for any number of buoyant vortices, and in any initial configuration.  This allows for a number of generalisations of the behaviour of buoyant vortex interactions.  It also has the advantage of fitting into the well-studied Hamiltonian framework.


The first study that considered two interacting buoyant point vortices was that of \citet{turn1960}.  Motivated by the release of effluent from chimneys, and buoyant `thermals' found in convection flows, he investigated the motion of two vortices of opposite circulation and equal buoyancy anomaly relative to the ambient fluid.  The motion of this buoyant vortex \emph{couple}, was found by \citet{turn1960} for the special configuration of a purely vertical trajectory in the direction of the buoyancy force (i.e., for a buoyant couple rising, or dense couple falling).  In this case he made the non-intuitive finding that increasing the buoyancy difference leads to a decrease in the rate of rise of the buoyant couple. 

Since the original paper of \citet{turn1960}, there have been a number of studies that have extended his analysis, nearly always focussing on the important case of two vortices with opposite circulation \citep[e.g.,][]{ravi2017}, and often in more complicated ambient environments with shear and/or stratification \citep[e.g.,][]{saff1972,gart1998}.  {In a series of papers by Arendt \citep{arendt1993a,arendt1993b,arendt1996} the motion of buoyant vortices in polytropic stratified fluids, where the pressure and density are related through a power law, have been systematically explored.  Many of the results in \citet{arendt1996} will be found to have analogues in the present investigation where we focus on buoyant vortices in a homogeneous ambient.}  Recently, \citet{ravi2017} examined the collision and collapse of such buoyant vortex couples, and conducted simulations using many vortex patches to demonstrate the presence of this mechanism in a random field of vortices. 

{Similar to the general framework established by \citet{arendt1996}, we formulate the motion of many buoyant point vortices of arbitrary buoyancy anomaly and circulation in a homogeneous fluid, into a single general framework.}  Our motivation for such a formulation arises from the study of instability development in stratified shear layers, which has been found to lead to the formation of concentrated patches of buoyancy and vorticity \citep[e.g.,][]{carp2010,smyt2012}.  An idealised model for the interaction of these buoyant vortex structures, such as formulated here, is therefore of interest in order to understand the evolution of stratified shear layers, with implications for oceanic and atmospheric mixing \citep{fern1991}.

The paper is organised as follows.  In the next section we derive a series of conservation equations for the motion of buoyant vortices, and demonstrate that they are captured by a Hamiltonian framework.  Given this general formulation of the laws governing the motion of buoyant point vortices, we then go on to explore the solutions by increasing the number of vortices from $N = 1$, to $N = 3$, in sections \ref{sec:one_vortex} to \ref{sec:three_vortices}.  Special cases of interest, such as the buoyant vortex couple, and an exploration of chaotic motion in the buoyant 3-vortex problem, will be examined along the way.  A summary and conclusions follows in the final section.  

\section{General formulation} \label{sec:formulation}


\subsection{Conservation laws} \label{subsec:consn_laws}

In this subsection \ref{subsec:consn_laws}, we present a brief background of the formulation of vortex conservation laws for Boussinesq flows \citep[for details about homogeneous flows see][]{saff1992}.  

Consider an inviscid, non-diffusive, Boussinesq fluid inside a simply connected domain of \emph{fixed} volume $V$ (which could be infinite). The background state, with density $\rho_a$, is assumed to be irrotational and in hydrostatic balance. Denoting perturbation velocity, vorticity, pressure  and density respectively by $\mathbf{u}$, $\boldsymbol{\omega} (\equiv \nabla \times \textbf{u})$, $p$ and $\rho$, the perturbed Boussinesq Euler equations on integration yield
\begin{equation}
\frac{\partial}{\partial t}\int_{V}\mathbf{u}(\mathbi{x},t) \md V=\int_{V}\bigg( \mathbf{u}\times\boldsymbol{\omega}+\mathbi{F} \bigg) \md V - \int_{\partial V}\bigg(\frac{p}{\rho_a}+\frac{1}{2}|\textbf{u}|^{2}\bigg)\hat{\mathbf{n}}\, \md A,
\label{eq:BNS}
\end{equation}
where 
$$\mathbi{F}\equiv-\frac{g}{\rho_a} \rho\hat{\mathbf{j}}.$$
Here $g$ denotes gravity, $\hat{\mathbf{j}}$ the unit vector in the vertical $y$ direction, and $\hat{\mathbf{n}}$ the unit vector normal to the bounding surface $\partial V$.
We will focus on the dynamics of an isolated vortex patch (i.e., vorticity distribution with a compact support) of volume $V^{(v)} \in V$. 
The vortex patch may have a density different from $\rho_a$, hence yielding the body force (or buoyancy) term $\mathbi{F}$. The vector quantity $\mathbf{u}\times\boldsymbol{\omega}$ is referred to by different names by different authors, such as the \emph{vortex force} \citep{saff1992}, \emph{Magnus force} \citep{ravi2017}, or the \emph{lift force} \citep{ligh1986}, however, we will use the latter term in this study.  

By taking $\mathbi{x} \times (\nabla \times )$ of the integrands in  Eq.\, (\ref{eq:BNS}) and making use of various vector calculus identities, detailed in \citet{saff1992}, we arrive at 
\begin{equation}
   \oder{}{t}\mathbf{I}^{(v)}=\int_{V^{(v)}}\bigg( \mathbf{u}_\mathrm{ext}\times\boldsymbol{\omega}+\mathbi{F} \bigg) \md V .
\label{eq:linimp_iso} 
\end{equation}
This expresses the conservation of \emph{linear impulse} of the vortex patch, defined as
$$\mathbf{I}^{(v)}=\frac{1}{2}\int_{V^{(v)}}\mathbi{x}\times\boldsymbol{\omega}\, \md V .$$ As discussed in \cite{saff1992} and \cite{dav2015}, this result relies on the fact that both $\mathbf{u}$ and $p$ due to an isolated vortex decay as $\mathcal{O}(r^{-3})$, where $r$ is the distance from the vortex.  Therefore, the surface integrals which appear in this process vanish.  In Eq.\, (\ref{eq:linimp_iso}), the quantity $\mathbf{u}_\mathrm{ext}$ represents the induced velocity within $V^{(v)}$ due to all vorticity external to it.  Similarly, defining the \emph{angular impulse} of an isolated vortex as
$$\mathbf{L}^{(v)}=-\frac{1}{2}\int_{V^{(v)}}\norm{\mathbi{x}}^2\boldsymbol{\omega}\,  \md V,$$
we obtain the conservation of angular impulse
 \begin{equation}
\oder{}{t}\mathbf{L}^{(v)}=\int_{V^{(v)}}\mathbi{x}\times\bigg( \mathbf{u}_\mathrm{ext}\times\boldsymbol{\omega}+\mathbi{F} \bigg) \md V.
\label{eq:ang_imp_iso}
\end{equation}

If multiple such vortex patches are present (and are far away from the domain boundary) in the irrotational background flow field, $\mathbf{u}_\mathrm{ext}$ on a given patch would arise from the action of the other vortex patches present in the system. It can be obtained  using the Biot-Savart law, which inverts the vorticity field $\boldsymbol{\omega}_\mathrm{ext}$ existing due to the other patches:
\begin{equation}
\mathbf{u}_\mathrm{ext} (\mathbi{x})=\frac{1}{2(n-1)\pi}\int_{V}\frac{\boldsymbol{\omega}_\mathrm{ext}(\mathbi{x}^\prime)\times (\mathbi{x}-\mathbi{x}^\prime)}{\norm{\mathbi{x}-\mathbi{x}^\prime}^n} \md V^\prime,
    \label{eq:biot_sav}
\end{equation}
where {the number of spatial dimensions is $n=2$ or $3$}.
From Eq.\, (\ref{eq:linimp_iso}) and Eq.\, (\ref{eq:ang_imp_iso}), we recover the well-known result that in the absence of any external velocity field {and applied force}, both linear and angular impulses of a homogeneous vortex patch (i.e., with the same density as the background) are constants of motion.

In this paper, we will be focusing on two-dimensional ($x$--$y$) systems. The vorticity vector is then given by $\boldsymbol{\omega}=-\omega\mathbf{\hat{k}}$, where $\omega(x,y) = \partial u/\partial y - \partial v/\partial x$, and $(u,v)$ are fluid velocities in the $(x,y)$ directions.  The linear and angular impulses of an isolated vortex patch of area $A^{(v)}$ are given by \citep{saff1992} 
\begin{equation}
    \mathbf{I}^{(v)}=-\int_{A^{(v)}}\omega\mathbi{x}\times\hat{\mathbf{k}}\, \md A\,\,\,\,\mathrm{and}\,\,\,\, \mathbf{L}^{(v)}=\frac{1}{2}\int_{A^{(v)}}\norm{\mathbi{x}}^2\omega\hat{\mathbf{k}}\, \md A.
    \label{eq:2d_impulses}
\end{equation}
The missing $1/2$ in $\mathbf{I}$ for two-dimensional flows is a consequence of the vortex lines not being closed.  {When working with two-dimensional flows we also consider area integrals to be expressed per unit length.}

The conservation of linear and angular impulse essentially represent rewritten forms for the linear and angular momentum that are more practical for vortex dynamics studies.  This will become apparent in the next subsection when we use them to formulate point vortex motions.

\subsection{Equations governing buoyant point vortex dynamics}

Consider $N$ vortices, denoted by the subscript $i=1,...,N$, that move across the unbounded $x$--$y$ plane in time, $t$.  We assume that the diameter of the vortices is much smaller than the distance between them, and so we can take them to be represented by `point vortices' at the locations $\mathbi{x}_i = (x_i,y_i)$.  The vorticity field is zero everywhere except at the locations of the vortices, where it has delta-function behaviour, i.e.,
\begin{equation} \label{omega}
\omega(x,y) = \sum_{i=1}^N \Gamma_i \delta (\mathbi{x} - \mathbi{x}_i) .
\end{equation}
Here $\Gamma_i$ indicates the circulation strength of vortex $i$, with a $\Gamma_i > 0$ indicating clockwise circulation.

In addition, we allow each vortex to be composed of fluid { with a density anomaly from the constant ambient density $\rho_a$}.  Each vortex will then experience a buoyant {point} force, and the field of these forces can be written as
\begin{equation} \label{F}
\mathbi{F}(x,y) = \sum_{i=1}^N \gt_i \delta (\mathbi{x} - \mathbi{x}_i) \hat{\mathbf{j}}.
\end{equation}
$\mathbi{F}$ is expressed as a force per unit {mass (per unit length in two dimensions)}, so that the `reduced gravity' of each vortex, $\gt_i$, is defined by the difference in density with the ambient {$\rho(x,y)$, and the vortex cross sectional area $A_i^{(v)}$, as 
\begin{equation} 
\tilde{g}_i \equiv \int_{A_i^{(v)}} \mathbi{F} \cdot \hat{\mathbf{j}} \: \mathrm{d}A = -\int_{A_i^{(v)}} \frac{\rho(x,y)}{\rho_a}g \: \mathrm{d}A .
\end{equation}
Hence $\gt_i>0$ implies that the vortex is (positively) buoyant.  This definition of the reduced gravity, $\gt_i$, differs from the usual definition by the area factor.  Since we are assuming the fluid to be Boussinesq, the inertia of the vortices resulting from their difference in density from the ambient fluid is neglected.  Note that in taking our vortex to be concentrated at a point and shrinking the area to zero, we ensure that the net buoyant force on the vortex patch is fixed. This limit might appear to violate the Boussinesq approximation, as the density difference with the ambient should become infinite, however, it must be kept in mind that these limits are only mathematical tools used when making a point approximation to a finite-sized vortex.  Another option would be to take $g \to \infty$, as in \citet{aref1989}, which would avoid this issue.  We observe that the integral terms in Eqs. (\ref{eq:linimp_iso})--(\ref{eq:ang_imp_iso}) preserve the conservation of linear and angular impulse in this point vortex representation \citep[e.g.,][]{turn1960}.}

As in the case of a system of homogeneous vortices \citep{kirc1876}, in the system just described it is possible to use a Hamiltonian framework to find the vortex trajectories.  The Hamiltonian $\h$, is given by the energy 
\begin{equation} \label{eq:hamiltonian}
\h = - \sum_{i,j}^N \frac{\Gamma_i \Gamma_j}{4 \pi} \mathrm{ln}[(x_i-x_j)^2+(y_i-y_j)^2] - \sum_{i=1}^N  \gt_i y_i , 
\end{equation}
being composed of both a `kinetic energy of interaction' (given by the first term on the right hand side), and a potential energy (last term), and is a constant of motion \footnote{In this Hamiltonian system the generalised coordinates, and momenta take the form of $q_i = \Gamma_i y_i$, and $p_i = x_i$, respectively, where $dp_i/dt = -\partial \h / \partial q_i$ and $dq_i/dt = \partial \h / \partial p_i$.}.  Note that the sum in the kinetic energy term is not carried out over the singular terms $i=j$. Substituting Eqs.\, (\ref{omega})--(\ref{F})  in the linear  impulse conservation equation Eq.\, (\ref{eq:linimp_iso}), and using the definitions in Eqs.\, (\ref{eq:biot_sav}, \ref{eq:2d_impulses}), we obtain the following Hamiltonian dynamical system of $2N$ dimensions:
\begin{subequations} \label{mom}
\begin{equation} \label{y_mom}
\Gamma_i \oder{x_i}{t} = -\frac{\partial \h}{\partial y_i} \quad \Rightarrow \quad \oder{}{t}I_y^{(i)} = \sum_{i,j}^N \frac{\Gamma_i \Gamma_j}{2 \pi}\frac{y_i-y_j}{(x_i-x_j)^2+(y_i-y_j)^2} +  \gt_i,
\end{equation}
\begin{equation}\label{x_mom}
\Gamma_i \oder{y_i}{t} = \frac{\partial \h}{\partial x_i} \quad \Rightarrow \quad -\oder{}{t}I_x^{(i)}=-\sum_{i,j}^N \frac{\Gamma_i \Gamma_j}{2 \pi}\frac{x_i-x_j}{(x_i-x_j)^2+(y_i-y_j)^2},
\end{equation}
\end{subequations}
where $\mathbf{I}^{(i)}=\left(I_x^{(i)},I_y^{(i)}\right)$ represents the linear impulse of vortex $i$. It is important to note that the summation terms in Eq.\, (\ref{mom}) arise from the externally induced lift force $\int_{V^{(v)}}\mathbf{u}_\mathrm{ext} \times \boldsymbol{\omega} \md V$ of the linear impulse conservation Eq.\, (\ref{eq:linimp_iso}).

 It is also possible to find global conservation laws for all buoyant vortices in a straight forward manner directly from the Hamiltonian description. In this regard we follow the analyses of \cite{lamb1895}, who deduced the conservation laws for homogeneous point vortices,
\begin{subequations} \label{mom1}
\begin{equation} \label{eq:x_mom1}
\sum_{i=1}^N \pder{\h}{x_i} = 0 \quad \Rightarrow \quad -\oder{}{t}I_x \equiv \sum_{i=1}^N \Gamma_i \oder{y_i}{t} =0 ,
\end{equation}
\begin{equation} \label{eq:y_mom1}
-\sum_{i=1}^N \frac{\partial \h}{\partial y_i} = \sum_{i=1}^N \gt_i \quad \Rightarrow \quad \oder{}{t}I_y \equiv \sum_{i=1}^N \Gamma_i \oder{x_i}{t} =  \sum_{i=1}^N \gt_i.
\end{equation}
\end{subequations}
Also, if we transform into polar coordinates, with $(x_i,y_i) = (r_i \cos \theta_i , r_i \sin \theta_i )$, it is possible to derive the conservation of the global angular impulse:
\begin{equation} \label{ang}
\sum_i \frac{\partial \h}{\partial \theta_i} = - \sum_{i=1}^N \tilde{g}_i x_i \quad \Rightarrow \quad \frac{\dd}{\dd t}L^2 \equiv \frac{1}{2} \frac{\dd}{\dd t} \sum_{i=1}^N \Gamma_i (x_i^2 + y_i^2) = -\sum_{i=1}^N \gt_i x_i. 
\end{equation}
This shows that when a global budget (including all vortices) is considered, the lift forces can be regarded as internal forces, and will cancel to produce no net effect.  

{The Hamiltonian in Eq.\,(\ref{eq:hamiltonian}) has correctly reproduced the conservation laws of linear and angular impulse, derived in a general form in the previous subsection, for a buoyant point vortex system.  This can be verified by substitution of Eqs.\,(\ref{omega})--(\ref{F}) into Eqs.\,(\ref{eq:linimp_iso})--(\ref{eq:ang_imp_iso}).  These conservation laws, and their alterations under the force of buoyancy from the homogeneous vortex case, will be critical for determining the nature of the solutions in what follows.  Note that the Hamiltonian itself is also a conserved quantity, corresponding to energy conservation.}  {These conservation laws and Hamiltonian structure also have an analogue in the configuration studied by \citet{arendt1996}.}


\section{A single buoyant vortex} \label{sec:one_vortex}

For the case of a single buoyant vortex (whose parameters are not denoted with a subscript)  the Hamiltonian equations of motion in Eqs.\, (\ref{y_mom})--(\ref{x_mom}) become
\begin{equation} \label{eq:single_vortex}
\dot{x} = \gt/\Gamma \quad \mathrm{and} \quad  \dot{y} = 0,
\end{equation}
where the dot represents ordinary differentiation with respect to $t$.  This shows that the vortex propagates in the horizontal direction with speed $U = \gt/\Gamma$, and merely expresses the conservation of linear impulse for a single vortex.  Note that for a vanishing buoyancy, this solution reduces to the homogeneous result of a stationary vortex. 
The non-intuitive result that a strictly vertical buoyancy force results in the horizontal motion of the vortex can be understood as follows.  The Hamiltonian formulation reveals that total energy $\h$ is a constant of motion.  Since there is no kinetic energy of interaction between vortices, the total potential energy must remain constant, giving $\dot{y}=0$.  However, in order to keep the vortex at the same vertical level, there must be an additional force that is needed to balance the vertical buoyancy force.  As long as the vortex is moving horizontally, this force is provided by the aerodynamic lift (or vortex force) that is present on any body possessing a circulation in a cross flow \citep{ligh1986}.  In terms of force per unit length, the lift force is $\rho_a \Gamma U$, and the buoyancy force is $\rho_a \gt$, which leads directly to the resulting translation velocity.  The appearance of an equivalent horizontal drift of a vortex once subject to a body force is also derived by \citet{saff1992}, {and discussed by \citet{arendt1993b}.} In Appendix \ref{sec:App1}, we numerically show that a buoyant point vortex (modeled as a Gaussian vortex patch) indeed travels horizontally with a constant speed.

One may also be tempted to infer that the horizontal motion of the buoyant vortex arises due to the absence of an inertial term {(from making the Boussinesq approximation)} that can lead to a vertical acceleration of the vortex under the action of gravity.  This can easily be added in an \emph{ad hoc} manner, and the equations of motion written as
\begin{equation}
\mu \ddot{x} + \Gamma \dot{y} =0 \quad \mathrm{and} \quad \mu \ddot{y} - \Gamma \dot{x} + \gt = 0 ,
\end{equation}
with $\mu$ a measure of the mass of the vortex.  A similar balance was considered in \citet{ravi2017}.  However, their model is based on potential flow around a circular cylinder and it is not clear how it may be constructed from the equations of motion  applied to vortex motions (i.e. proceeding similar to that outlined in section \ref{sec:formulation}). 
Proceeding nonetheless, a general solution for the velocities $(\dot{x},\dot{y})$ of this system can be found as  
\begin{equation}
\dot{x}(t) = C_1\cos(\alpha t) + C_2\sin(\alpha t) + \gt/\Gamma \quad \mathrm{and} \quad \dot{y}(t) = C_1\sin(\alpha t) - C_2\cos(\alpha t) ,
\end{equation}
where $C_1,C_2$ are constants that depend on the initial conditions.  This solution describes a horizontal drifting motion of the vortex at the speed $\gt/\Gamma$ and an oscillation at the frequency $\alpha \equiv \Gamma / \mu$.  

The vortex motion is comprised of a balance between inertia and the buoyancy and lift forces.  Note that if inertia is neglected, the lift-buoyancy balance leads to the constant horizontal drift of $\dot{x} = \gt/\Gamma$.  This arises from the fact that the lift force is always perpendicular to the vortex motion, and is the only force that can balance the vertical buoyancy force.  The oscillation arises due to the exchange between the kinetic and potential energy components of the total energy, which must be conserved.  The addition of inertia to the system, therefore does not significantly change the fundamental force balance, or the vortex trajectory.  This neglect of the vortex inertia will be appropriate as long as the dimensionless number 
\begin{equation}
\Pi \equiv \frac{\mu^2 \gt^4}{\Gamma^8} ,
\end{equation}
which quantifies the relative importance of the inertial term, is small.  If we approximate the mass of the vortex by its area, $A^{(v)}$, assume that the vortex is circular with radius $a$, and approximate the circulation by a solid body rotation with maximum velocity $V_\mathrm{max}$, we can write
\begin{equation}
\Pi^{1/4} \sim \frac{g' a}{V_\mathrm{max}^2} ,
\end{equation}
with $g' \equiv \Delta \rho g/\rho_a$ the standard reduced gravity.  This shows that $\Pi^{1/4}$ has the form of a Richardson number, or $\Pi^{-1/8}$ a Froude number.

\section{Two interacting buoyant vortices}

\subsection{Solution}
\begin{figure}
\begin{center}
\includegraphics[width=0.47\textwidth]{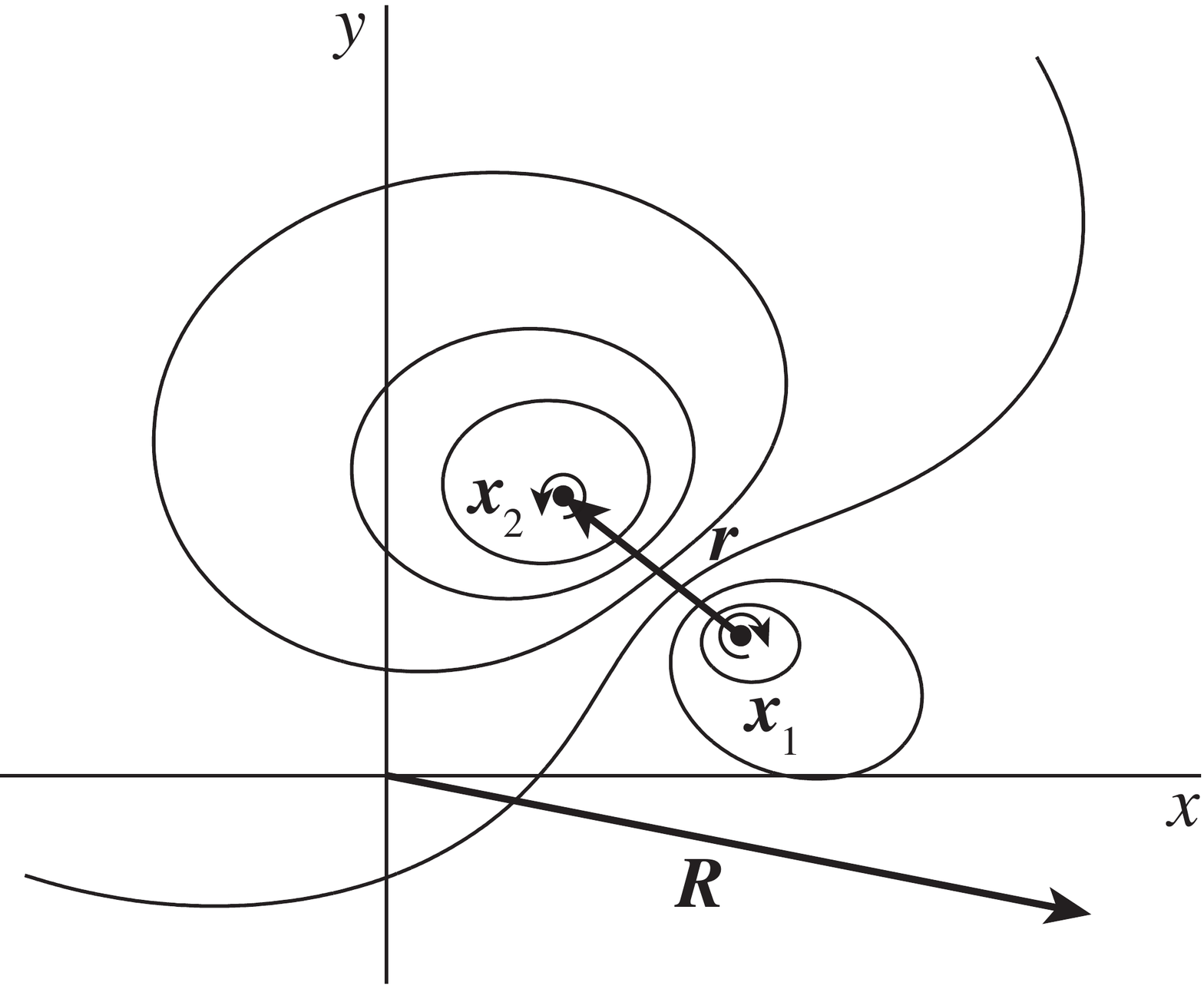}
\caption{Streamlines and notation for two interacting vortices.  In this case, a vortex couple with unequal circulation strength ($\Gamma_1 \neq -\Gamma_2$) is sketched.}
\label{basic}
\end{center}
\end{figure}

Using the Hamiltonian system Eqs.\, (\ref{y_mom})--(\ref{x_mom}) with $N=2$ vortices, now leads to a non-zero kinetic energy of interaction term, and considerably richer dynamics.  In analogy with the two-body problem of classical physics, it is possible to make considerable analytical progress in solving this system by introducing new variables given by the vortex separation
\begin{equation}
\mathbi{r} \equiv (\zeta,\eta) = (x_1-x_2, y_1-y_2) ,
\end{equation}
and the centre of vorticity
\begin{equation}
\mathbi{R} \equiv (X,Y) = \Big( \frac{\Gamma_1 x_1 + \Gamma_2 x_2}{\Gamma} , \frac{\Gamma_1 y_1 + \Gamma_2 y_2}{\Gamma} \Big) ,
\end{equation}
as illustrated in Fig~\ref{basic}.  
Here we have defined $\Gamma \equiv \Gamma_1 + \Gamma_2$, and the original coordinates can be recovered by
\begin{equation}
\mathbi{x}_1 = \mathbi{R} + \frac{\Gamma_2}{\Gamma} \mathbi{r} \quad \mathrm{and} \quad \mathbi{x}_2 = \mathbi{R} - \frac{\Gamma_1}{\Gamma} \mathbi{r} .
\end{equation}
With the new coordinates the Hamiltonian takes the form
\begin{equation}
\h = -\frac{\Gamma_1 \Gamma_2}{4\pi} \mathrm{ln}(\zeta^2 + \eta^2) - \frac{\gt_1 \Gamma_2 - \gt_2 \Gamma_1}{\Gamma}\eta - \gt Y ,
\end{equation}
where $\gt \equiv \gt_1 + \gt_2$. The original $2N$-dimensional Hamiltonian system has been reduced to a $(2N-2)$-dimensional Hamiltonian system, and we can rephrase the whole problem as the following set of equations:
\begin{subequations}
\begin{equation}
\frac{\Gamma_1 \Gamma_2}{\Gamma} \dot{\zeta} = -\frac{\partial \h}{\partial \eta}, \quad \frac{\Gamma_1 \Gamma_2}{\Gamma} \dot{\eta} = \frac{\partial \h}{\partial \zeta},
\end{equation}
for $\mathbi{r}$, and 
\begin{equation} \label{R_equations}
\Gamma \dot{X} = -\frac{\partial \h}{\partial Y}, \quad \Gamma \dot{Y} = \frac{\partial \h}{\partial X} ,
\end{equation}
\end{subequations}
for $\mathbi{R}$.  Notice that in this set of equations the motion of the centre of vorticity $\mathbi{R}$ is \emph{decoupled} from the vortex separation $\mathbi{r}$.  In particular, the centre of vorticity moves with the constant horizontal drift as if it were a single vortex with circulation $\Gamma$, and reduced gravity $\gt$.  This may be seen from Eq.\, \eqref{R_equations}, i.e., 
\begin{equation} \label{eq:dR}
\frac{\dd}{\dd t} \mathbi{R} = (\gt/\Gamma , 0) ,
\end{equation}
giving the same result as Eq.\, (\ref{eq:single_vortex}), which can easily be integrated.  Then we are left with a system of two first-order ODEs for the components of $\mathbi{r}$.  

At this point it is useful to non-dimensionalize the equations using $\Gamma^2/\gt$ as a length scale, and $\gt/\Gamma$ as a velocity scale.  The corresponding form for the Hamiltonian becomes
\begin{equation}
\h_\ast = -\frac{\Gamma_1 \Gamma_2}{\Gamma^2} \Big[ \frac{1}{4\pi} \mathrm{ln}(\zeta^2_\ast + \eta^2_\ast) + D \eta_\ast \Big] - Y_\ast ,
\end{equation}
where all variables with an asterisk are non-dimensional, and we have defined the dimensionless number $D \equiv (\Gamma / \gt)(\gt_1/\Gamma_1 - \gt_2/\Gamma_2)$.  The system of equations describing the trajectories is then written as
\begin{equation}
\dot{\zeta}_\ast = \frac{1}{2\pi} \frac{\eta_\ast}{\zeta^2_\ast + \eta^2_\ast} + D \quad \mathrm{and} \quad \dot{\eta}_\ast = -\frac{1}{2\pi} \frac{\zeta_\ast}{\zeta^2_\ast + \eta^2_\ast} .
\label{eq:2Ddynsys}
\end{equation}
Note that for $D=0$ the equations describing two homogeneous point vortices are recovered.  We can gain considerable insight into the solution for the vortex trajectories by using the fact that the total energy is conserved.  Since $\h$ is a constant of the motion, given by its initial value $\h(t=0) \equiv E$, it allows us to write 
\begin{equation} \label{trajectory}
\zeta^2_\ast + \eta^2_\ast = C^2 \exp(-4\pi D\eta_\ast)
\end{equation}
with 
\begin{equation}
C^2 \equiv \exp \Big[ -\frac{4\pi}{\Gamma_1 \Gamma_2} ( E + \gt Y ) \Big] 
\end{equation}
a dimensionless constant given by the initial conditions.  Note that from Eq.\, (\ref{eq:dR}) we can always define our coordinate system so that $Y=0$ for all $t$ (this is just a choice of datum for the potential energy).  

\begin{figure}
\begin{center}
\includegraphics[width=0.92\textwidth]{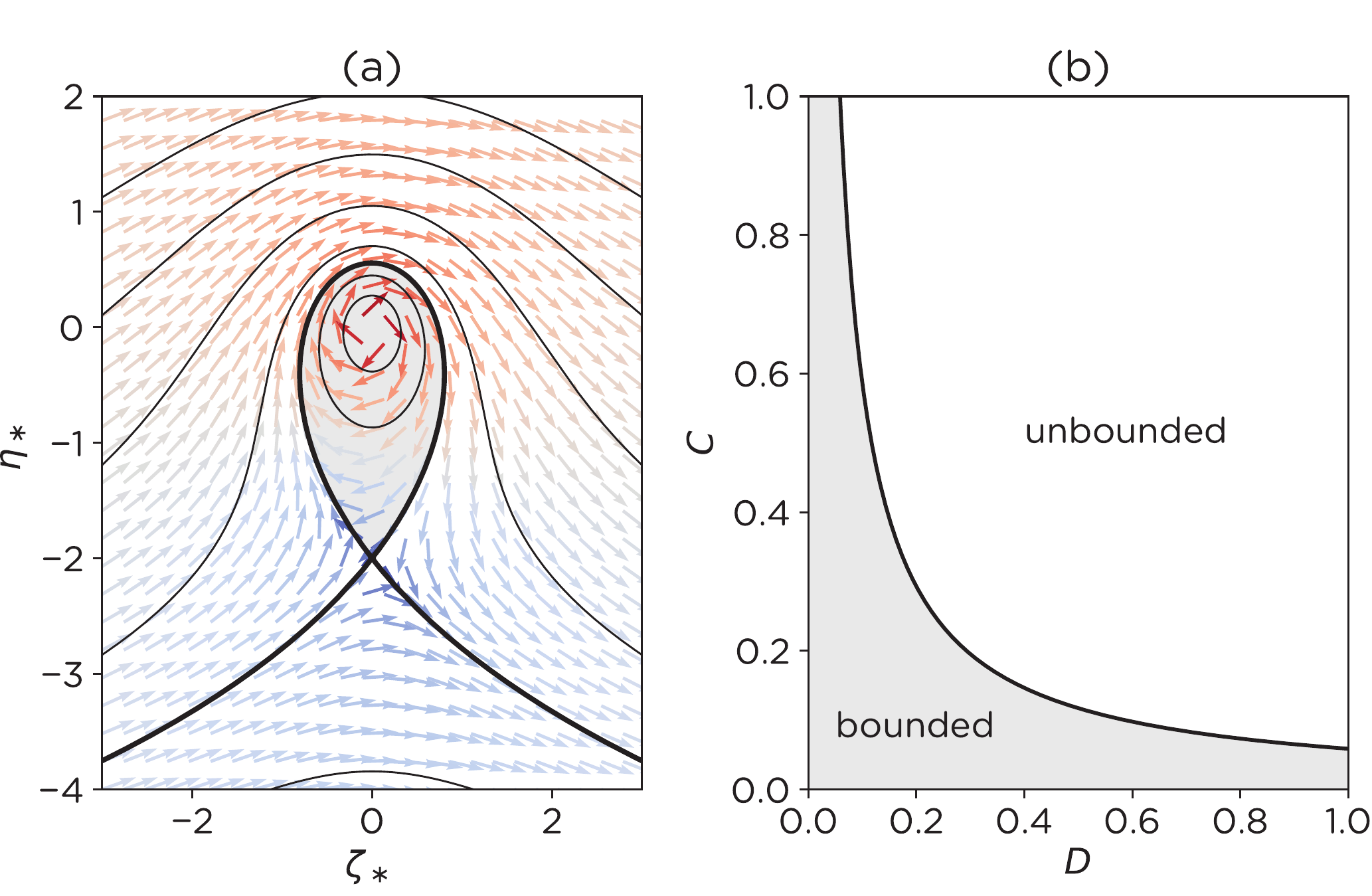}
\caption{(a) Phase portrait for the case of $D=(4 \pi)^{-1}$.  The contours represent values of constant $C^2$, with the thick contour representing the separatrix. It corresponds to a critical value of $C$ below which the orbits are bounded to each other, shown by the grey region.  The arrows denote the direction of the trajectories in phase space with the colours representing the speed of the point through phase space with red representing fast trajectories, and blue to slower trajectories.  (b) Critical curve in the $CD$-plane separating bounded and unbounded orbits.   }
\label{curves}
\end{center}
\end{figure}

Equation \eqref{trajectory} defines a set of curves that describe the vortex separation distance over the entire evolution of the vortex interaction, and are shown in Fig~\ref{curves}(a).  
{Similar to the results of \citet{arendt1996} for the case of a stratified polytropic fluid,} we can generally split the solution into two different categories: (i) those that remain bounded (i.e., within a finite distance) to the centre of vorticity, and (ii) those that are unbounded.  For a given value of $D$, there is a critical value of the dimensionless parameter $C$, given by $C_{\mathrm{cr}} = (2\pi e D)^{-1}$ and shown in Fig~\ref{curves}(b) (which can be thought of as a measure of the initial distance between the vortices, or equivalently as the initial energy), above which there are unbounded orbits (Fig~\ref{curves}, white area). The separatrix, represented by the thick dark contour in Fig~\ref{curves}(a), separates the bounded and unbounded regions. This separatrix appears as a homoclinic orbit, with its hyperbolic fixed point located at
$$ (\zeta^F_\ast,\eta^F_\ast)=\bigg(0,-\frac{1}{2\pi D}\bigg),$$
and has eigenvalues $\pm 2\pi D^2$ and eigenvectors $(\mp 1, 1)$. Hence the eigenvalues and eigenvectors are independent of the sign of $D$, leading to a clockwise circulation (in the bounded region) for both $D>0$ and $D<0$. The far-field flow, however, depends on the sign of $D$ as per Eq.\, (\ref{eq:2Ddynsys}). We also note in passing that $(0,0)$ is \emph{not} a fixed point, but on the contrary it corresponds to a point of infinite relative velocities, as can be seen from the coloured arrows in Fig~\ref{curves}(a).  This singularity is not realistic as we have assumed that the separation of the vortices is much larger than their diameter.

If the initial energy of the vortices is large enough, i.e., $C$ is large enough, then unbounded orbits will result, and this is more likely to occur for large $D$.  Note that for the singular case of homogeneous point vortices $D=0$, and we recover the result that all vortex trajectories are bounded.  The dependence of the orbits on the parameter $D$ can be interpreted more easily if we write it in terms of the individual vortex velocities assuming there is no interaction between them, i.e., $U_i \equiv \gt_i/\Gamma_i$.  Then we have
\begin{equation}
D = \frac{U_1-U_2}{U_\mathrm{cv}} ,
\end{equation}
where $U_\mathrm{cv} \equiv \gt/\Gamma$ is the velocity of the centre of vorticity.  As we would expect intuitively, when the difference in the individual vortex velocities is large compared to that of the centre of vorticity, they are able to `escape' from a bounded orbit.

\subsection{Special cases}

\subsubsection{Bounded and unbounded orbits - an example}

It is helpful to examine an example of the transition between bounded and unbounded vortex trajectories.  Fixing $D=(4\pi)^{-1}$ and choosing $C=2e^{-1} \pm \epsilon$ will produce a bounded trajectory for $C< C_\mathrm{cr} = 2e^{-1}$, and an unbounded trajectory for $C>C_\mathrm{cr}$.  In practise, we choose $\Gamma_1 = 2\Gamma_2$, and can compute the trajectories relative to $\mathbi{R}$.  
\begin{figure}
\begin{center}
\includegraphics[width=0.99\textwidth]{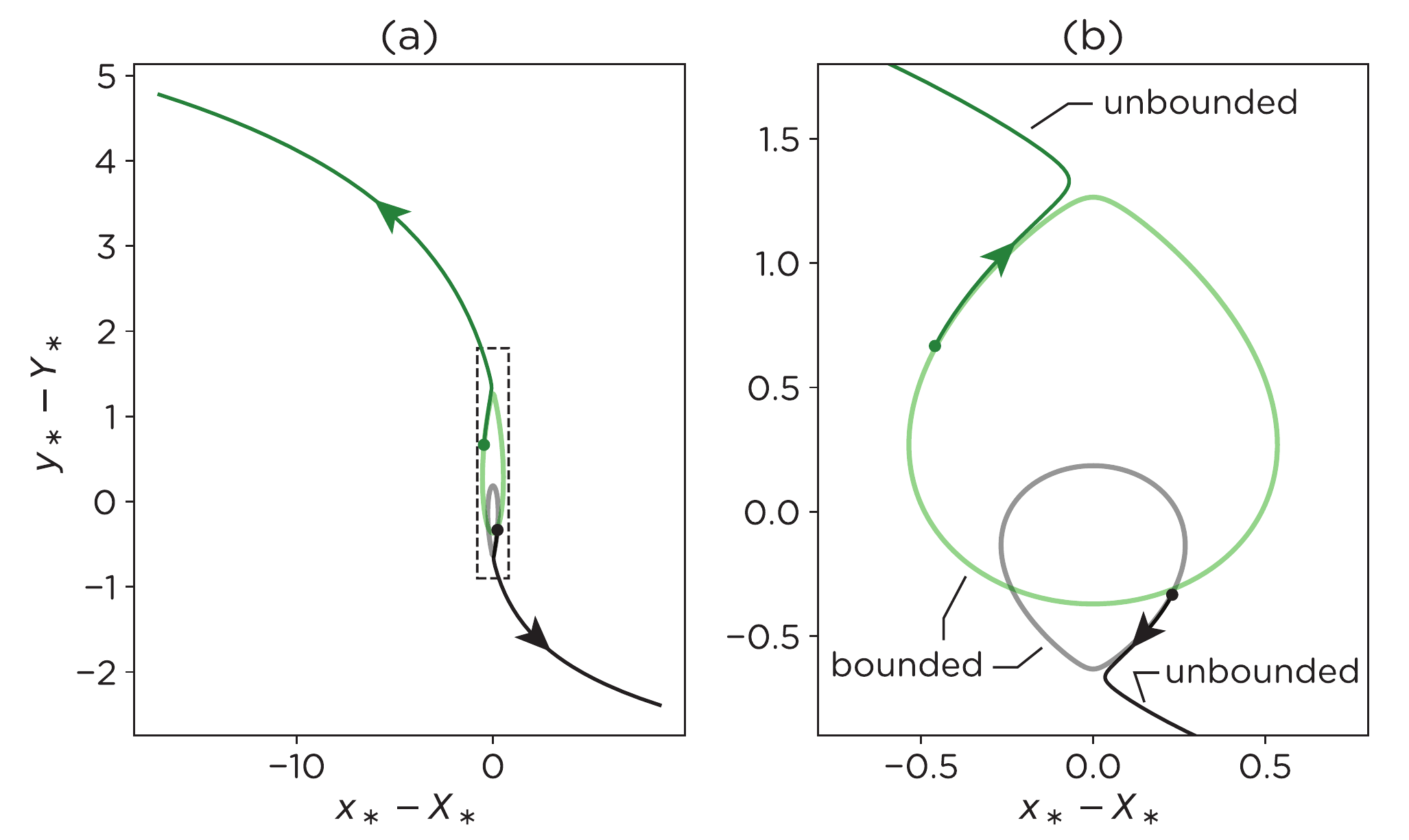}
\caption{Vortex trajectories in a frame of reference moving with the centre of vorticity for two buoyant vortices with bounded and unbounded orbits.  We have used $C=2e^{-1} \pm \epsilon$ to transition between the two cases, with $\epsilon = 0.001$, and the minus (plus) sign for the (un)bounded case.  In both plots $\Gamma_1 = 2\Gamma_2$ and $D=(4\pi)^{-1}$.  A close up of the dashed rectangular region in (a) with the bounded orbits is shown in (b).  The dark line corresponds to vortex 1, the green line to vortex 2, with dots indicating the vortex starting positions.}
\label{traject1}
\end{center}
\end{figure}

The results are shown in Fig~\ref{traject1}.  In the unbounded case, the vortices approach a constant horizontal trajectory in opposite directions, as the distance between them increases, and thus the interaction weakens.  Recall that for a vanishing interaction each vortex will approach a horizontal trajectory at the speed $\gt_i/\Gamma_i$.  Only an extremely small change in the initial energy (and therefore position) of the vortices, expressed by $C$, alters the orbits of the two from bounded to unbounded (Fig~\ref{traject1}b).  This diverging of the orbits occurs as the two are farthest from one another, and the interaction is weakest.  This sensitive dependence on the initial condition in the two vortex problem is only present at this point in parameter space close to the critical separation between bounded and unbounded orbits.

\subsubsection{The vortex couple}

In the original analysis of \citet{turn1960}, he considered the special case of the strictly vertical propagation of a buoyant vortex couple, i.e., vortices with $\Gamma_1 = -\Gamma_2$.  This vortex configuration often arises in idealized models of convection `thermals', and in starting jets.  In these applications, and in Turner's analysis, it is assumed that the couple is formed from an initial buoyant cloud, or jet, and therefore it is also the case that $\gt_1 = \gt_2 \equiv \gt_0$.  In this subsection we extend Turner's analysis and examine the motion of buoyant vortex couples in general.  

To solve for the trajectory of the vortex couple we cannot use the formulation in the previous section since $\Gamma = 0$; instead we will use the conservation laws directly.  The conservation of linear impulse Eq.\, (\ref{mom}) reduces to 
\begin{equation} \label{eq:r}
\frac{\dd}{\dd t} \mathbi{r} = (2\gt_0/\Gamma_0,0)
\end{equation}
where $\Gamma_0 \equiv \Gamma_1 = -\Gamma_2$.  This equation states that in order for a vortex couple to conserve linear impulse the vertical distance between vortices remains constant in time, while the horizontal distance changes linearly with time.  The total energy for the couple can be written 
\begin{equation}
    \label{eq:h_couple}
\h = \frac{\Gamma_0^2}{2\pi}\ln |\mathbi{r}| - 2\gt_0 \bar{y} ,
\end{equation}
where we have defined the mean vertical position of the couple as $\bar{y} \equiv (y_1 + y_2)/2$.  Differentiating this equation with respect to $t$, and combining it with Eq.\, (\ref{eq:r}) allows us to solve for the evolution of the couple.  

For large times, we can effectively ignore the initial separation of the vortices and, by Eq.\, (\ref{eq:r}), write 
\begin{equation}
    \zeta(t) \sim \frac{2\gt_0}{\Gamma_0}t.
\end{equation}
This can be combined with Eq.\, (\ref{eq:h_couple}) to find the long-time evolution of the elevation of the couple as,
\begin{equation}
\oder{\bar{y}}{t} \sim \frac{\Gamma_0^2}{4 \pi \tilde{g}_0} t^{-1} .
\end{equation}
This relation demonstrates a number of noteworthy results.  First, buoyant couples will rise, and dense couples will fall, inline with our intuition.  Second, they will do so at a decreasing rate, proportional to $t^{-1}$.  Last, it is interesting that as the magnitude of the buoyant force (measured by $\gt_0$) is increased, the rate of elevation change \emph{decreases}.  This is a generalisation of Turner's result that less buoyant thermals will rise faster.  The reason for this is clearly that the rate of separation increases with increasing $|\gt_0|$, due to the dominant balance for large times between the lift and buoyancy forces on the individual vortices.  This results from increasing horizontal motion of the vortices required to increase the lift in response to increased buoyancy.

\begin{figure}
\begin{center}
\includegraphics[width=0.99\textwidth]{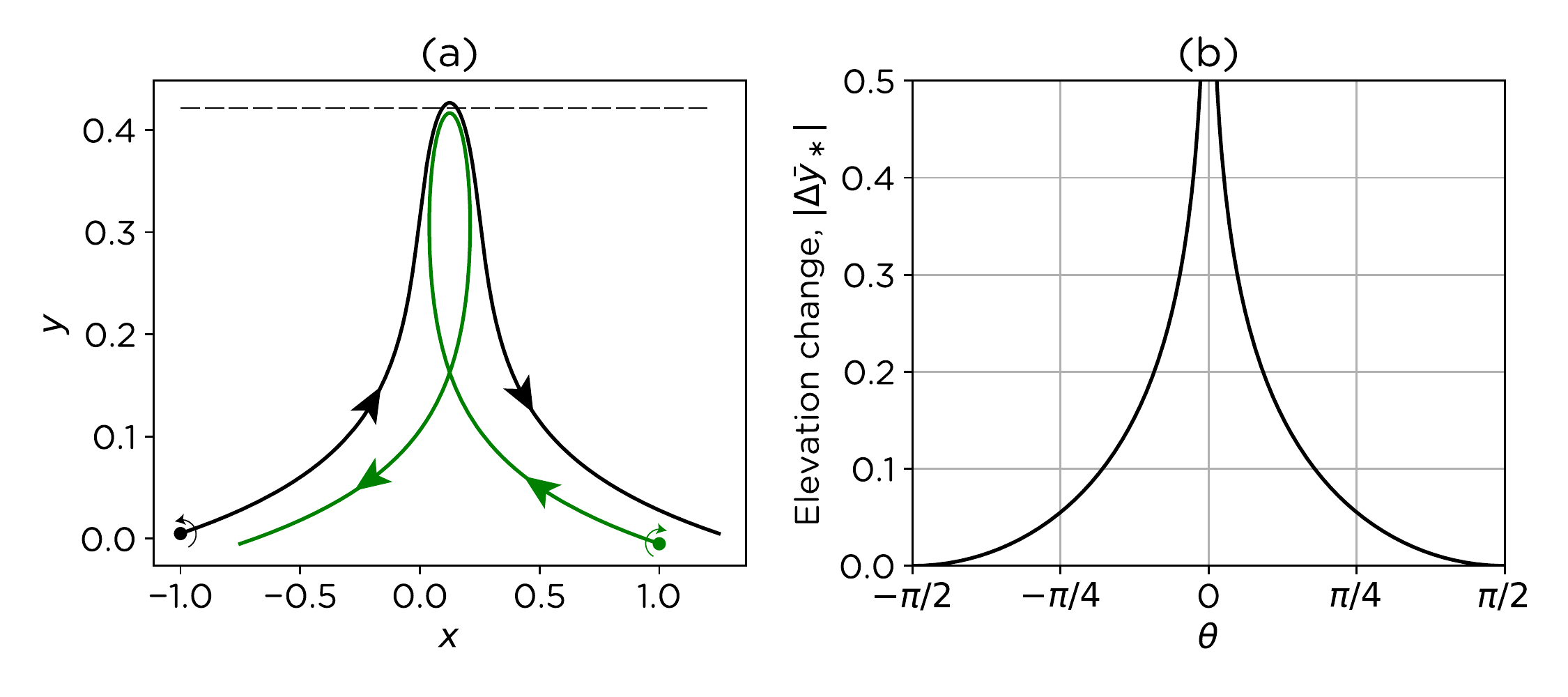}
\caption{(a) Trajectories of a vortex couple with negative (downward-directed) buoyancy, and a circulation oriented as shown.  The horizontal dashed line indicates the elevation change of the couple, $\Delta \bar{y} = 0.42$.  Parameters are given by $\Gamma_0 = -1$, $\gt_0 = -1$, $\zeta_0 = -2$, and $\eta_0 = 0.01$, with the dark line corresponding to vortex 1, the green line to vortex 2, and dots indicating the vortex starting positions.  (b) Maximum elevation change in the vortex couples (made dimensionless through $\Delta \bar{y}_\ast \equiv \Delta \bar{y} \gt_0/\Gamma_0^2$) as a function of the vortex orientation angle, $\theta$.  The singularity at $\theta = 0$ represents the case of purely vertical orientation of the couple, as discussed in the text.}
\label{f:couple}
\end{center}
\end{figure}

For short times, however, the couples can propagate in a direction that is opposite to the buoyant forces.  An example in Fig~\ref{f:couple} shows such an orientation and trajectory.  In an arbitrary initial orientation it can be shown that the rate of elevation change of the couple is
\begin{equation}
    \oder{\bar{y}}{t} = \frac{\Gamma_0}{2\pi} \frac{\zeta}{|\mathbi{r}|^2},
\end{equation}
where $|\mathbi{r}|^2 = \zeta(t)^2 + \eta_0^2$ with $\eta_0$ the initial vertical separation (which is constant in time).  This shows that it is the vertical advection speed of each vortex on the other that is responsible for the mean rate of elevation change (but with the separation governed by the conservation of linear impulse).  Using the conservation of $\h$ allows us to derive a maximum vertical displacement of the couple against buoyancy forces of
\begin{equation}
\Delta \bar{y} = \frac{\Gamma_0^2}{4\pi \gt_0} \ln \Big( \frac{|\eta_0|}{|\mathbi{r}_0|} \Big) ,
\end{equation}
where $|\mathbi{r}_0|$ is the initial separation of the couple.  This can alternatively be expressed in terms of the angle, $\theta$, that a line connecting the two vortices makes with the horizontal,
\begin{equation}
\Delta \bar{y} = \frac{\Gamma_0^2}{4\pi \gt_0} \ln (|\sin \theta|) .
\end{equation}
This relationship for the dimensionless elevation change, $\Delta \bar{y}\tilde{g}_0/\Gamma_0^2$, is plotted in Fig.\, \ref{f:couple}(b).

Note that the special case treated by Turner (with $\eta_0 =0$ and $\theta = 0$) can result in a singularity in finite time when the couple is directed opposite to the buoyant force, as it results in the collision of the vortices and an infinite vertical velocity of the couple.  This result is clearly unrealistic as our assumption that the diameter of the vortices is much smaller than the distance between them breaks down.  This case has been examined by \citet{ravi2017} who found that the collision results in significant deformation of the vorticity field into complicated structures, leading to the collapse of the individual vortex cores.  The process of vortex couple collapse is also likely to happen for small, but non-zero $\theta$, as the distance between the vortices decreases during the elevation change of the couple (Fig.\, \ref{f:couple}a).  This will likely lead to a decrease in the elevation change predicted by the curve in Fig.\, \ref{f:couple}(b).

%

\subsubsection{A `pseudo-homogeneous' case}

In this case, we note that trajectories of the buoyant vortices can occur as if they were homogeneous when $D=0$.  This occurs when $\gt_1 \Gamma_2 = \gt_2 \Gamma_1$, or similarly if the individual vortex speeds are equal, i.e., $U_1=U_2$.  The solution for $|\mathbi{r}|$ is
\begin{equation}
\zeta^2 + \eta^2 = C^2
\end{equation}
which corresponds exactly to the equations for the homogeneous two-vortex problem with $C$ the initial distance between the vortices.  
The only difference between the trajectories of these buoyant vortices and two similar homogeneous vortices is the constant motion of the centre of vorticity at the speed $U_\mathrm{cv} = \gt/\Gamma$.  The vortices are also rotating about the centre of vorticity at the frequency of $\Gamma/2\pi C^2$, independent of buoyancy.  Rather than plotting the trajectories in the $xy$-plane, it is also possible to get an accurate qualitative feel for the trajectories by using the $\zeta \eta$-plane.  This can be seen by writing 
\begin{equation}
\mathbi{x}_1 - \mathbi{R} = \frac{\Gamma_2}{\Gamma} \mathbi{r} \quad \mathrm{and} \quad \mathbi{x}_2 - \mathbi{R} = -\frac{\Gamma_1}{\Gamma} \mathbi{r} ,
\end{equation}
so that the individual vortex trajectories, in a frame of reference moving with the centre of vorticity, are just a scaled version of the curve in the $\zeta \eta$-plane.



\section{Three buoyant vortices} \label{sec:three_vortices}

A general property of Hamiltonian systems is the appearance of chaotic motions as the number of degrees of freedom is increased \citep{tabor}.  In the case of homogeneous point vortices, it is well known that chaotic motions appear once the number of vortices (equal to the number of degrees of freedom of the Hamiltonian system) $N \geqslant 4$ \citep{aref1982,aref1983}.  The key to integrating the equations, and thereby eliminating the possibility of chaotic trajectories, is to determine $N$ independent integrals (constants) of motion, $F_1, ..., F_N$, where $F_j=\mathrm{constant}$ for all vortex positions and all $t$.  
These integrals of the motion are usually formed from the conservation laws that we found in Eq.\, \eqref{mom} and Eq.\, \eqref{ang}, as well as the Hamiltonian (energy) itself.  For example, in the homogeneous vortex case, three integrals of motion can be identified as $\h$, $I_x^2 + I_y^2$, and $L^2$ \citep{aref1982}.  These integrals can be used to effectively reduce the number of degrees of freedom so that the system may be solved (in theory) by an integration \footnote{In Hamiltonian systems, each integral allows the elimination of a \emph{pair} of conjugate variables (e.g., $\Gamma_i y_i,\,x_i$ in our case). Thus if we have $N$ integrals of motion for a system of $2N$ Hamiltonian equations (i.e. $N$ degrees for freedom, and $N$ vortices) we can, therefore, reduce the problem to quadratures. Such systems are said to be Liouville integrable.}.  {Although techniques exist to derive integrals of motion, such as Lie algebra and symmetries of the Legrangian \citep[e.g.,][]{chapman1978,bolsinov1999,badin_book}, one often relies on intuition \citep[see][p. 39]{tabor}.}


For the global conservation laws of the buoyant vortex system derived in  Eqs.\, (\ref{mom1})-(\ref{ang}) we can see that the addition of buoyancy forces has resulted in non-constant angular impulse $L$, and vertical linear impulse, $I_y$.  We are therefore left with only two conserved quantities $\h$, the energy, and $I_x$, the horizontal linear impulse. Hence, the buoyant $3$-vortex problem appears to be non-integrable, {as was also the case found for the stratified polytropic fluid system of \citet{arendt1996}.}  The purpose of this section is therefore to answer the question: \emph{Can three interacting buoyant vortices exhibit chaotic motion?}

\begin{figure}
\begin{center}
\includegraphics[width=0.85\textwidth]{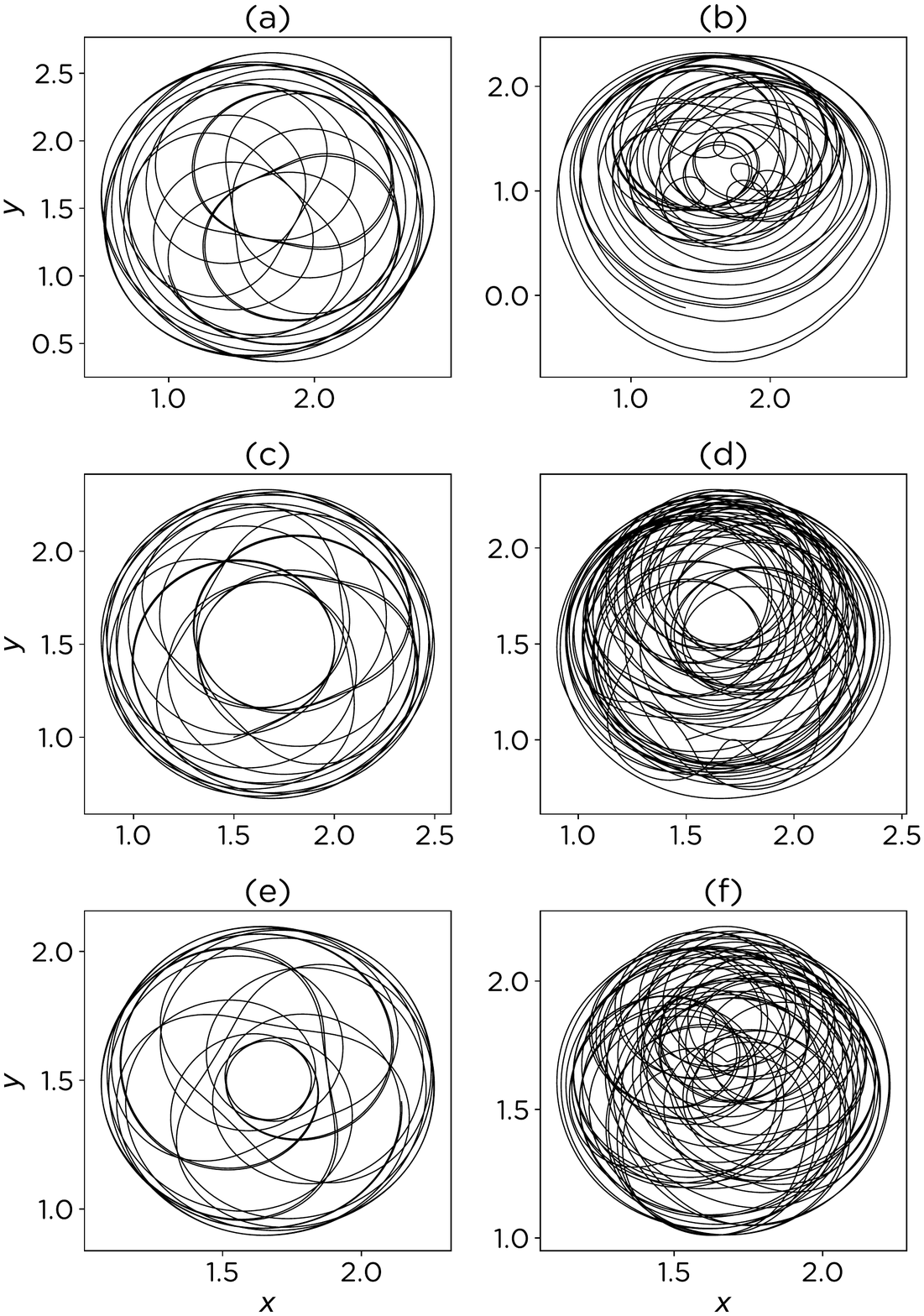}
\caption{Vortex trajectories for the homogeneous (a,c,e) and buoyant (b,d,f) 3-vortex problems.  Each row shows vortices with the same starting positions [i.e., $(1,1)$, $(1.5,1)$ and $(2,2)$ respectively for each row, starting from top] and same circulations ($\Gamma_i = \{1,2,3\}$, starting from top), with the vortices in (b,d,f) having $\gt_i = \{0.02,0,-0.02\}$, respectively.  }
\label{f:3-vortex}
\end{center}
\end{figure}

To address this question, we begin by showing in Fig.\, \ref{f:3-vortex} an example of the evolution of three buoyant vortices, and comparing to the homogeneous case (which is known to be non-chaotic).  In both cases, the circulation of the vortices is identical with $\Gamma_i = \{1,2,3\}$, but the buoyant 3-vortex system has $\gt_i = \{-0.2,0,0.2\}$ (Fig.\, \ref{f:3-vortex}b,d,f).  Note that  since $\sum_i \gt_i = 0$, by Eq.\, (\ref{eq:y_mom1}), there is no net movement of the centre of vorticity.  It can be seen from Fig.\, \ref{f:3-vortex} that, whereas the homogeneous vortices trace out regular symmetric patterns, the buoyant vortices have irregular asymmetric trajectories.  This is an indication of the presence of chaos in the buoyant 3-vortex problem, as we might expect from the loss of an independent integral of motion.

\begin{figure}
\begin{center}
\includegraphics[width=0.8\textwidth]{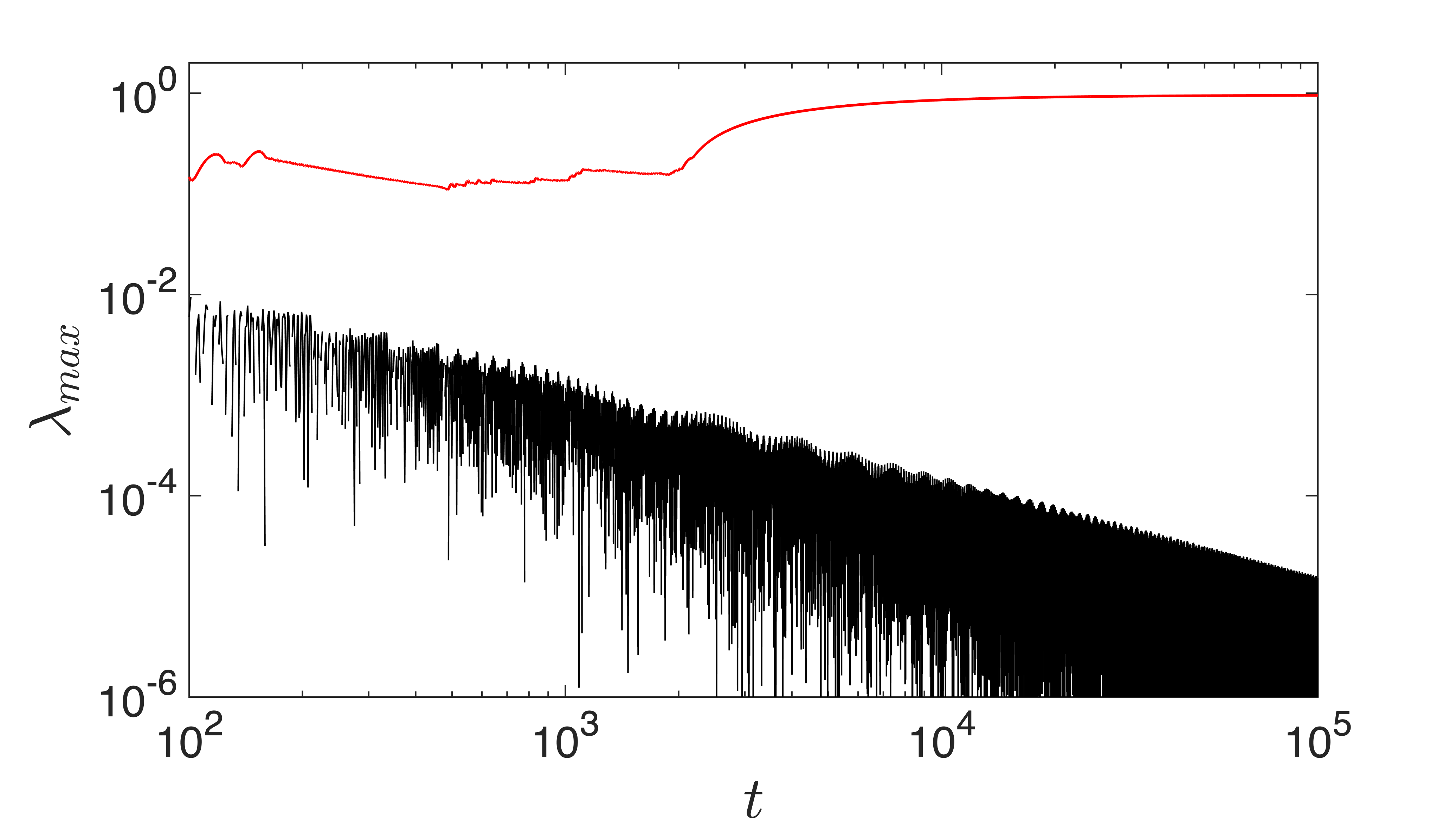}
\caption{Maximum Lyapunov exponents for three homogeneous vortices (black curve) and three buoyant vortices (red curve). Circulation and initial position of the vortices are the same in both cases, corresponding to the systems plotted in Fig.\, \ref{f:3-vortex}.}
\label{f:lyap}
\end{center}
\end{figure}

To further address the presence of chaos in the buoyant 3-vortex problem, we have numerically calculated the Lyapunov exponents. A defining feature of chaos, is that nearby initial conditions will produce diverging solutions after finite times.  This feature is quantified through the maximum Lyapunov exponent defined as
\[
\lambda_\mathrm{max} \equiv \lim_{t\rightarrow\infty}\lim_{\delta\mathbf{X}_{0}\rightarrow0}\,\,\frac{1}{t}\ln\left(\frac{|\delta\mathbf{X}(t)|}{|\delta\mathbf{X}_{0}|}\right).
\]
It characterises the exponential rate of separation of infinitesimally close trajectories whose initial separation is  $\delta\mathbf{X}_{0}$.  An autonomous non-linear dynamical system with $\lambda_\mathrm{max}>0$ is non-integrable, implying chaos is a possibility.  We have numerically computed the Lyapunov exponents up to $t= 100,000$ using the procedure outlined in \citet{wolf1985}. In Fig.\, \ref{f:lyap}, $\lambda_\mathrm{max}$ is shown for both three homogeneous vortices (black curve) and three buoyant vortices (red curve), corresponding to the case examined in Fig.\, \ref{f:3-vortex}. In the absence of reduced gravity, i.e., for the homogeneous 3-vortex case, $\lambda_\mathrm{max} \rightarrow 0$ with increasing $t$, as expected. However, the presence of buoyancy in the problem leads to a positive $\lambda_\mathrm{max}$, indicating that two nearby trajectories will exponentially diverge, further supporting the finding that chaos is present. 

\begin{figure}
\begin{center}
\includegraphics[width=0.75\textwidth]{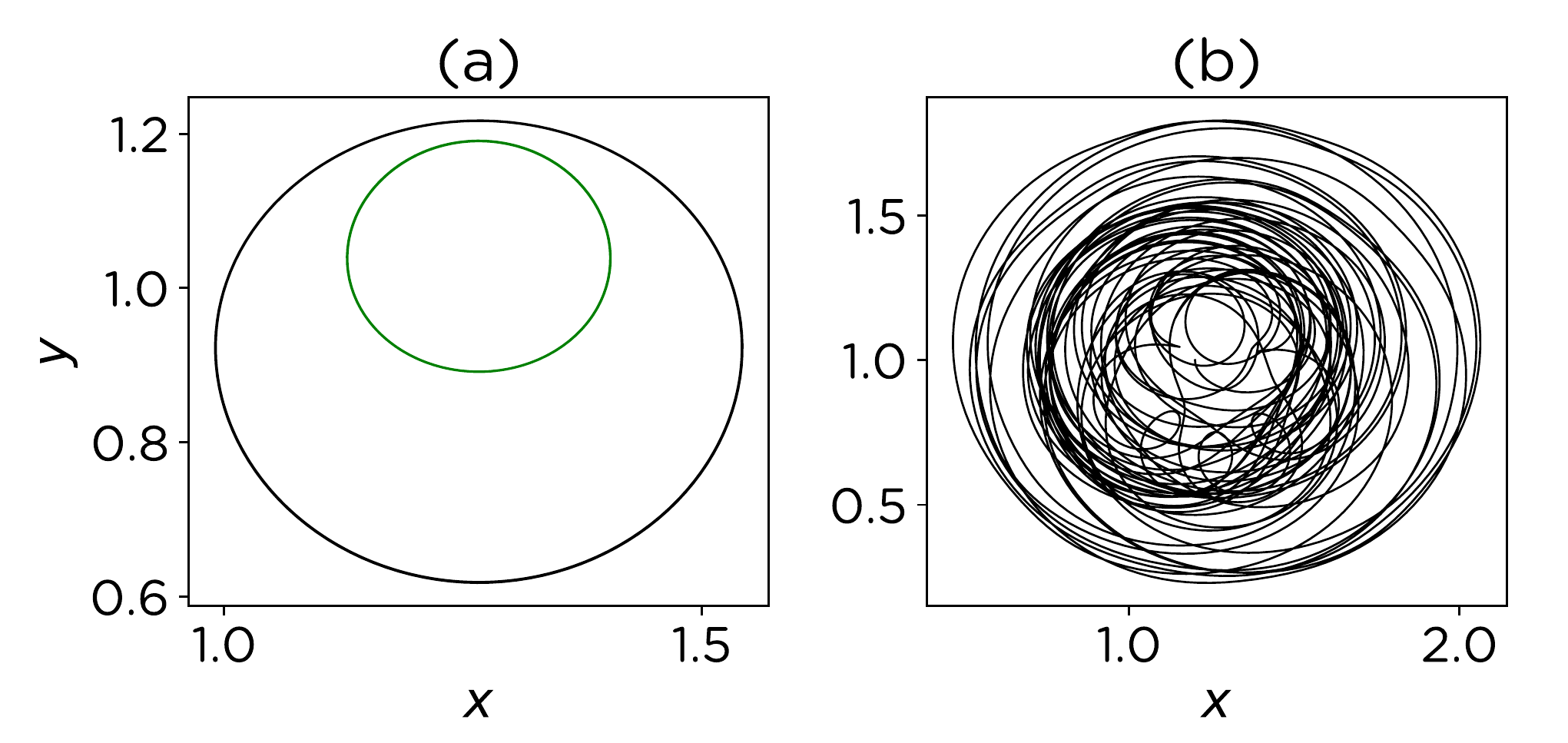}
\caption{Illustration of chaotic advection in the restricted buoyant 3-vortex problem.  (a) Trajectories of the two buoyant vortices with non-zero circulation (black and green curves), along with (b) the passive vortex (i.e., neutrally buoyant with vanishing circulation).}
\label{f:chaotic_advection}
\end{center}
\end{figure}

A point worth mentioning in the context of differences between the homogeneous and buoyant 3-vortex problems is the concept of \emph{chaotic advection} -- simple time-dependent flows can cause chaotic motion of tracer parcels.  The basic idea stems from the fact that the motion of a point vortex with vanishing circulation and zero buoyancy anomaly will behave as a tracer particle, being advected by the fields induced by the other vortices present. When such a `tracer vortex' is included in a field of three homogeneous point vortices, although the unsteady flow field produced by the three point vortices is integrable, the motion of the tracer vortex is not \citep{aref2007}. This result echos a similar finding for the 3-body problem with one of the point masses being negligible, and is commonly called the restricted 3-body problem.  This concept can be easily extended to understand chaotic advection due to buoyant vortices, with an example of the restricted buoyant 3-vortex problem shown in Fig.\, \ref{f:chaotic_advection}.  The flow field produced by two buoyant vortices is integrable (Fig.\, \ref{f:chaotic_advection}a) with repeated closed orbits, however, the motion of neutrally buoyant tracers in this flow field is not.  The chaotic advection of the tracer particle is demonstrated in Fig.\, \ref{f:chaotic_advection}(b) to consist of irregular non-repeating orbits indicative of chaos.  

\section{Summary and conclusions}     

In this paper we have formulated a general Hamiltonian framework for the investigation of buoyant point vortices {in a homogeneous ambient fluid}.  This extends previous work {in this area} that looked only at special cases of buoyant vortices of relevance for particular applications.  {It reveals analogous results to the general Hamiltonian framework developed for vortices in a stratified polytropic ambient fluid by \citet{arendt1996}.}  We have provided a systematic study of the 1- and 2-vortex problems, and discovered that the evolution of two buoyant vortices can be split into two different solution types: those in which the vortices remain bounded (i.e., within a finite distance) to each other, and those in which the vortices drift apart in time.  The boundary between these two cases was also derived.  As the number of vortices is increased to 3, the solutions become complex, and irregular chaotic motions result.  This feature of the buoyant 3-vortex problem is in contrast to the homogeneous 3-vortex problem, and arises due to the loss of symmetry in the Hamiltonian due to the buoyancy force, thus reducing the number of integral invariants. We also investigate the possibility of chaotic advection of tracer parcels arising from the simple, time dependent flow field induced by two buoyant vortices.  Future work could include a deeper investigation of chaotic motion in the 3-buoyant vortex problem, including a mapping of chaotic regions throughout parameter space, as well as investigating the statistical mechanics of many buoyant vortices.

\begin{acknowledgments}
We would like to acknowledge funding from the Helmholtz Foundation under the PACES II programme, and the Alexander von Humboldt Association for Humboldt research fellowship to A. Guha carried out at the Helmholtz-Zentrum Geesthacht.
\end{acknowledgments}

\appendix

\section{Numerical validation of the constant horizontal drift of a buoyant  vortex patch}
\label{sec:App1}

\begin{figure}
\begin{center}
\includegraphics[width=0.75\textwidth]{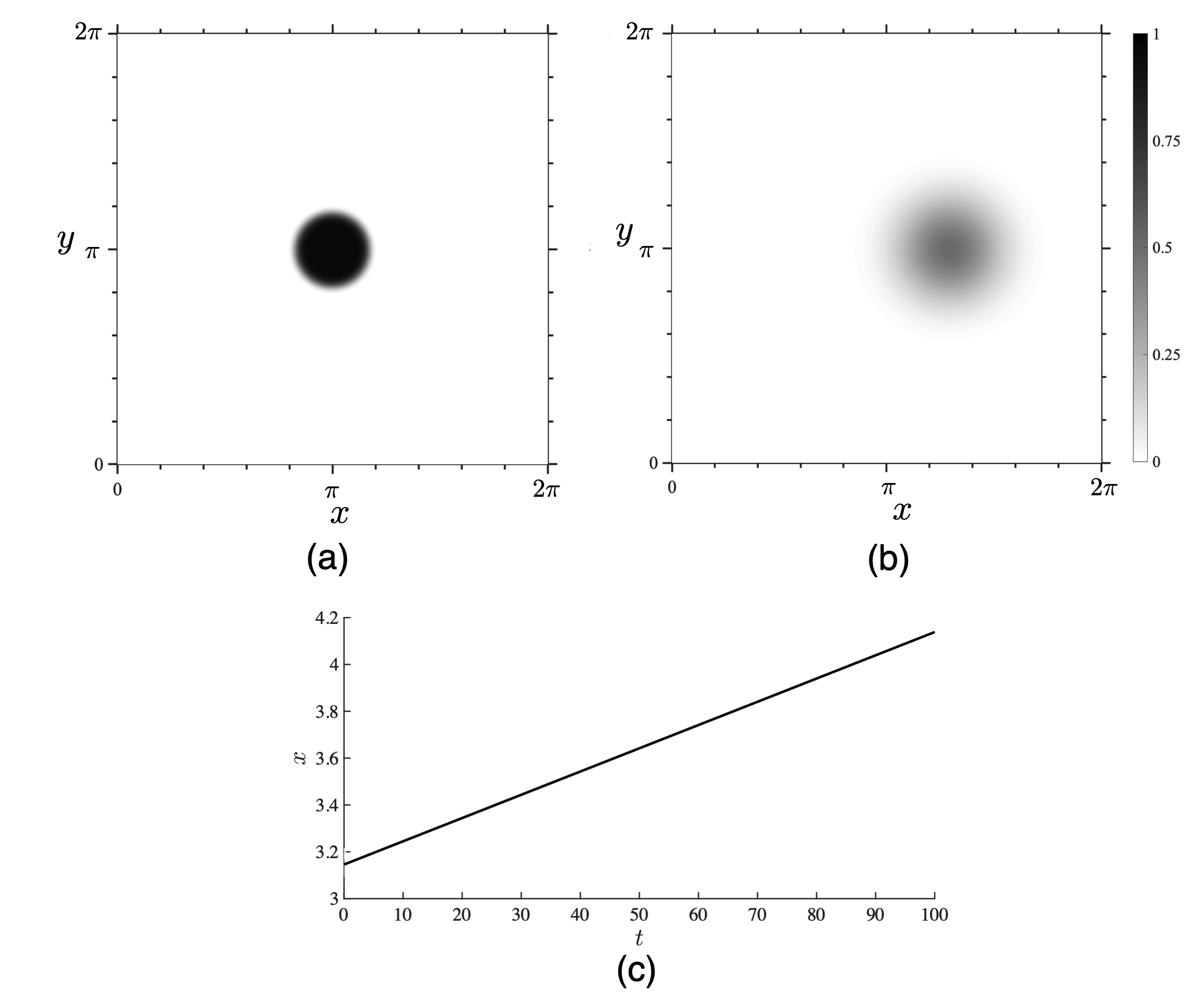}
\caption{Numerical simulation of the horizontal drift of a buoyant vortex patch: (a) at $t=0$ and (b) at $t=100$. (c) The abscissa of the centre of the vortex patch as a function of time. }
\label{f:val}
\end{center}
\end{figure}

We numerically solve the non-dimensional perturbation vorticity and  scalar (density) conservation equations in two dimensions ($x$-$y$):
\begin{align}
\frac{D\omega}{D t} & = -\dfrac{1}{\mathrm{Fr}^2}\dfrac{\partial \rho}{\partial x}+\dfrac{1}{\mathrm{Re}}\nabla^2\omega,\\ 
\frac{D\rho}{D t} & = \dfrac{1}{\mathrm{RePr}}\nabla^2\rho,
\end{align}
where $D/Dt$ denotes the material derivative, $\mathrm{Re}$, $\mathrm{Fr}$ and $\mathrm{Pr}$ respectively denote the Reynolds, Froude and Prandtl numbers. The above set of equations are solved using a pseudospectral code in a $2\pi\times2\pi$ domain with doubly periodic boundary conditions. The number of mesh-points used are $(N_x,N_y)=(512,512)$; the Crank–Nicolson method is used for time integration. The following parameters are assumed: $\mathrm{Re}=10^3$, $\mathrm{Fr}=10$ and $\mathrm{Pr}=1$. {We simulate a generalized Gaussian vortex centered at ($\pi,\pi$) having the following vorticity and density structure at $t=0$:
\begin{align}
\omega & =\exp\left[-\left\{((x-\pi)^2+(y-\pi)^2)/0.3\right\}^\beta\right],\\
\rho & =\exp\left[-\left\{((x-\pi)^2+(y-\pi)^2)/0.3\right\}^\beta\right],
\end{align}
where $\beta=6$. While $\beta=1$ results in a standard Gaussian distribution, higher values of $\beta$ allow a flatter peak, resulting in a vortex patch of nearly constant vorticity and density (i.e. approximating a `buoyant' Rankine vortex).}
We show the initial and final states of the vortex patch in Figs.\, \ref{f:val}(a) and \ref{f:val}(b), respectively. In Fig.\, \ref{f:val}(b), we find that the vortex has diffused, as expected in a viscous, diffusive flow. Nevertheless, the vortex indeed travels along a horizontal trajectory with a constant speed (Fig.\, \ref{f:val}c).  {This speed can be derived based on Eq.\,\eqref{eq:single_vortex} for a finite sized buoyant vortex patch giving $U = g'_0/\Omega_0$, where $\Omega_0$ is the near-constant initial vorticity of the patch and $g'_0$ the buoyancy anomaly (standard reduced gravity).  Non-dimensionalising this speed gives $U_\ast = \mathrm{Fr}^{-2} = 0.01$, corresponding to our chosen numerical setup.  This provides an excellent match to the results of the translation speed observed in the simulation of $U_\ast = 0.0099$.  }

\bibliographystyle{apalike}

\end{document}